# Deep Clustering via Center-Oriented Margin Free-Triplet Loss for Skin Lesion Detection in Highly Imbalanced Datasets

Şaban Öztürk, Tolga Çukur, *Senior Member, IEEE*

*Abstract*— Melanoma is a fatal skin cancer that is curable and has dramatically increasing survival rate when diagnosed at early stages. Learning-based methods hold significant promise for the detection of melanoma from dermoscopic images. However, since melanoma is a rare disease, existing databases of skin lesions predominantly contain highly imbalanced numbers of benign versus malignant samples. In turn, this imbalance introduces substantial bias in classification models due to the statistical dominance of the majority class. To address this issue, we introduce a deep clustering approach based on the latent-space embedding of dermoscopic images. Clustering is achieved using a novel center-oriented margin-free triplet loss (COM-Triplet) enforced on image embeddings from a convolutional neural network backbone. The proposed method aims to form maximally-separated cluster centers as opposed to minimizing classification error, so it is less sensitive to class imbalance. To avoid the need for labeled data, we further propose to implement COM-Triplet based on pseudo-labels generated by a Gaussian mixture model. Comprehensive experiments show that deep clustering with COM-Triplet loss outperforms clustering with triplet loss, and competing classifiers in both supervised and unsupervised settings.

Manuscript received XX.XX.XXXX; revised XX.XX.XXXX; accepted XX.XX.XXXX. Date of publication XX.XX.XXXX; date of current version XX.XX.XXXX. The work of T. Çukur was supported by TUBA GEBIP 2015, BAGEP 2017 awards. (Corresponding author: Şaban Öztürk)

Ş. Öztürk is with the Department of Electrical and Electronics Engineering, Amasya University, TR-05001 Amasya, Turkey, also with the Department of Electrical and Electronics Engineering, Bilkent University, TR-06800 Ankara, Turkey, also with the National Magnetic Resonance Research Center, Bilkent University, TR-06800 Ankara, Turkey, (e-mail: saban.ozturk@amasya.edu.tr).

T. Çukur is with the Department of Electrical and Electronics Engineering, Bilkent University, TR-06800 Ankara, Turkey, also with the National Magnetic Resonance Research Center, Bilkent University, TR-06800 Ankara, Turkey, and also with the Neuroscience Program, Sabuncu Brain Research Center, Bilkent University, TR-06800 Ankara, Turkey (e-mail: cukur@ee.bilkent.edu.tr).

## I. INTRODUCTION

SKIN cells that undergo a controlled development process under normal conditions divide abnormally to form masses in cancers. The prevalence of skin cancers has been steadily increasing in recent decades due to elevated exposure to harsh environmental factors and aging populations [1-3]. Early diagnosis is critical in improving the survival rate in deadly skin cancers such as melanoma. However, access to expert dermatologists might be limited for many patients, particularly in low-income countries [4]. Thus, automated screening based on dermoscopic images can improve detection rates and treatment outcomes across patient populations under risk [5]. Traditional methods for skin-lesion detection build classifiers based on hand-crafted features [6, 7]. Recent studies have instead adopted deep learning to learn data-driven features for improved accuracy and generalization [8-14]. The common approach in this domain is to leverage a convolutional neural network (CNN) with a softmax output layer to classify disease based on deep features of skin-lesion images [15, 16].

Learning-based classifiers for medical images ideally require large training datasets with balanced samples across different classes [17]. Unfortunately, this condition is difficult to meet in rare diseases such as melanoma, where skin-lesion samples are expected to be from a majority class of non-melanoma tissue [15]. For instance, popular public databases for melanoma typically have over two-orders-of-magnitude imbalance between malignant and benign samples. In turn, this imbalance can introduce unwanted biases in classification models that are trained to maximize overall detection accuracy, potentially elevating their false negative rates and limiting generalizability [18]. Therefore, there is a need for learning-based methods that alleviate biases in melanoma detection due to imbalance in skin-lesion datasets.

Several important approaches have been proposed to treat sample imbalance for learning-based classifiers in the literature. The first group of studies have leveraged data augmentation [16] or oversampling [19] methods to ensure that models are trained on a matching number of samples from each class. While these balancing methods are powerful when the rate of original data imbalance is moderate, their utility might be limited on skin-lesion datasets with substantial imbalance. In particular, repeated sampling of dermoscopic images from the minority class can increase risk of overfitting [20]. The second group of studies have instead adopted transfer learning or few-shot learning approaches [21] to pre-train networks in a different domain where balanced datasets are available. These methods avoid oversampling of the minority class since relatively compact datasets are often sufficient for fine-tuning of pre-trained models on skin-lesion datasets [22]. Yet, domain differences between pre-training and fine-tuning stages can introduce potential limitations in generalization performance.

Here, we introduce a deep clustering approach for melanoma detection from dermoscopy images to improve reliability against data imbalance in skin-lesion datasets. Unlike direct classifiers that optimize for detection accuracy, our approach maximizes as a proxy metric distance between cluster centers in a latent-embedding space that contains dense semantic information [23]. To learn discriminative embeddings, we introduce a novel COM-Triplet loss function for improved reliability in the identification of cluster centers over the traditional triplet loss. During inference, proximity to learned cluster centers in the embedding space is used for disease detection. To avoid the need for expensive class labels, we further introduce an unsupervised variant to compute the COM-Triplet loss where pseudo-labels are obtained via a Gaussian mixture model (GMM). Comparisons against competing methods and ablation studies are conducted to demonstrate both the supervised and unsupervised variants of the proposed method. Our results indicate cluster separation in a latent embedding space is a more resilient measure against data imbalance than detection accuracy in direct classifiers.

Our main contributions are summarized below:

● We introduce a deep-clustering method for melanoma detection that maximizes cluster separation in an embedding space to improve reliability against imbalanced training datasets.
● We introduce a novel COM-Triplet loss for learning discriminative embeddings that adaptively updates inter-cluster distance across the training procedure, unlike traditional triplet loss that maintains a fixed distance from the origin independently for positive and negative classes.
● We introduce an unsupervised variant of deep clustering based on pseudo-labels for embedded images generated via a GMM.

The rest of the paper is organized as follows; Section II provides a literature survey on skin-lesion classification; Section III provides the problem definition and details of the proposed method; Section IV contains experimental details; Section V presents results, while Section VI discusses the implications of our findings.

## II. RELATED WORK

Traditional studies on skin-lesion detection have mainly used low-level visual features directly related to color and morphology [6, 24], and hand-crafted mid-level features such as intuitive [25] or wavelet features [26]. Improved performance has been reported when using multiple different feature sets simultaneously [27, 28], albeit feature selection has been adopted to maintain low dimensionality in aggregated sets processed by classical machine-learning models [29, 30]. That said, traditional methods relying on hand-crafted features often show suboptimal performance with limited generalization under domain shifts.

Deep learning methods instead forego hand-crafted features in favor of a deep hierarchy of data-driven features. In the domain of skin lesions, a common approach rests on CNN models with softmax output layers to detect disease [31]. Recent studies have proposed numerous advances to improve classification accuracy of skin lesions. On the architectural front, advanced methods include wavelet domain CNN models [32, 33], synergic models that contain an ensemble of CNNs [34], multi-tasking models that leverage dermoscopy images along with their segmentation features [16], attention-gated CNN or self-attention transformer models [9, 35, 36]. On the algorithmic front, proposed techniques include domain transfer of pre-trained feature sets [37], augmentation via GAN-based synthetic sample generation [38, 39], and combination of multiple imaging modalities and patient metadata [40]. While these previous methods have enabled notable performance benefits in lesion classification, they do not explicitly consider high data imbalance between classes.

Common public datasets for skin-lesion classification typically show a high degree of imbalance between data samples from majority and minority classes. For instance, the ratio of the largest to the smallest class is 58.21 in the HAM10000 dataset [41], and 54.04 in the ISIC2019 dataset [42]. A number of recent studies on skin-lesion detection have focused on improving classification performance under such data imbalance via resampling procedures. Oversampling of the minority class has been reported to improve classification accuracy by balancing the training dataset [9, 43]. Other studies have employed standard or adversarial augmentation methods to increase the number of samples from the minority class [12, 13, 44]. While undersampling of the majority class is a viable alternative, there are mixed results in the literature regarding its utility in treating data imbalance [10, 11, 45]. Although resampling methods can help alleviate biases due to moderate levels of data imbalance, repeated sampling of few images from the minority class under heavy data imbalance can increase the risk of overfitting.

Alternative approaches for addressing data imbalance include transfer learning or loss weighting procedures. In transfer learning, classification models are pre-trained in a separate domain with limited class imbalance (e.g., ImageNet) and then fine-tuned on a compact skin-lesion dataset that can be undersampled to maintain inter-class balance [37, 46]. Although they alleviate the need for large training sets, transfer-learned models can show suboptimal generalization performance under substantial domain shifts between the pre-training and testing domains. In loss-term weighting, models are trained directly in the target application domain, albeit training loss is modified to give higher weight to errors in detecting the minority class [47, 48]. Prescribing a loss that emphasizes performance inversely with the relative proportion of minority-class samples can mitigate biases in a specific dataset, albeit this approach requires manual intervention and retraining when the rate of data imbalance changes across datasets.

## III. METHODOLOGY

### A. Direct Classifiers for Melanoma Detection

Given the lower incidence rate of melanoma compared to other skin lesions, it is highly challenging to collect large datasets with balanced samples across malignant versus benign tissue. For the binary problem of melanoma detection, this implies that the training set will contain a disproportionately large number of samples from the majority non-melanoma class ($C_{maj}$), albeit relatively few samples from the minority melanoma class ($C_{min}$). For instance, $C_{maj}$ and $C_{min}$ account for respectively 98.24% and 1.76% of the samples in the ISIC2020 skin-lesion dataset analyzed here (see Section IV.A for details). In turn, this gross data imbalance can introduce undesirable biases toward the majority class in common deep-learning classifiers trained to maximize overall detection accuracy, even when using weighted loss functions [18].

For the binary problem of melanoma detection, a direct classification model predicts a probability distribution over two classes given as input dermoscopic images. Let $D=\{X,Y\}^Z$ be a skin-lesion dataset with $Z$ samples where $X$ are dermoscopic images, $X \in R^{(256 \times 256) \times Z}$, and $Y$ are class labels, $Y \in [0,1]^Z$. The classifier is typically trained to *minimize* a cross-entropy loss:

$$L_{CE} = \frac{1}{Z}\sum_{i=1}^{Z}\left[-\left(w_{min}\left(Y_i \log(Y_i')\right) + w_{maj}\left((1-Y_i)\log(1-Y_i')\right)\right)\right] \quad (1)$$

where $Y'$ denotes the predicted probability for melanoma for the $i^{th}$ input image, and $w_{maj}$, $w_{min}$ stand for class weights. The standard approach is to weight the loss-term components for the two classes equally, while focusing on matching $Y$ and $Y'$ as closely as possible. However, since the majority class contains a substantially larger amount of samples, the classifier can trivially improve its performance by biasing predictions towards the majority class, thereby reducing detection sensitivity for the minority class. Non-equal weights in Eq. 1 can be used to partly alleviate such bias, but any differences between the training and test sets regarding the level of class imbalance can then introduce suboptimal performance [9].

### B. Deep Clustering for Melanoma Detection

In this study, we introduce a deep-clustering approach for melanoma detection to reduce training biases due to data imbalance. During training, the proposed approach first estimates cluster centers for melanoma and non-melanoma classes in a latent image-embedding space. To learn discriminative embeddings, a novel COM-Triplet loss function is used. During inference, proximity to learned cluster centers in the embedding space can be used for disease detection.

**Supervised Deep Clustering (SDC):** We first introduce a supervised variant of the proposed method for cases where a labeled training set is available. The proposed method leverages a CNN model to map dermoscopy images onto a latent space, $\Phi:X \rightarrow E$, where $\Phi$ is the mapping and $E$ are the resultant embeddings of dimensionality $S$, $E \in R^S$. To maintain discriminability in the embedding space, images belonging to the same class should be located in close proximity compared to images of opposing classes. A common approach for learning discriminative embeddings is based on the triplet loss that aims to maximize across-cluster over within-cluster distances [23, 49]. To do this, distances of an anchor image selected at random (*A*) with a positive image (*P*) from the same class, and with a negative image (*N*) belonging to the opposite class are computed:

$$L_{triplet} = \frac{1}{M}\sum_{i=1}^{M} \max\left\{0, d\left(E_A^i, E_P^i\right) - d\left(E_A^i, E_N^i\right) + \alpha\right\} \quad (2)$$

where $d$ is cosine distance and $\alpha$ is a constant margin value that serves as a lower threshold for cluster separation. During the *n*th iteration of the training process, a batch of *M* images are randomly selected for *A*, and depending on the class of each sample of *A*, labels in the training dataset are used to guide the selection of respective samples for *P* and *N*. In this case, the embeddings of these components are expressed as $E_{\{A,P,N\}} \in R^{S \times M}$. Because the triplet loss aims to maintain shorter *A-P* distance than *A-N* distance, it is possible that *P* remains equidistant to *A* and *N*, or even closer to *N* than *A* (Fig. 1a). Furthermore, prescription of a manually selected margin with constant value can degrade clustering performance due to suboptimal margin selection for a given dataset, and due to native variations in cluster separation across training iterations (Fig. 1c).

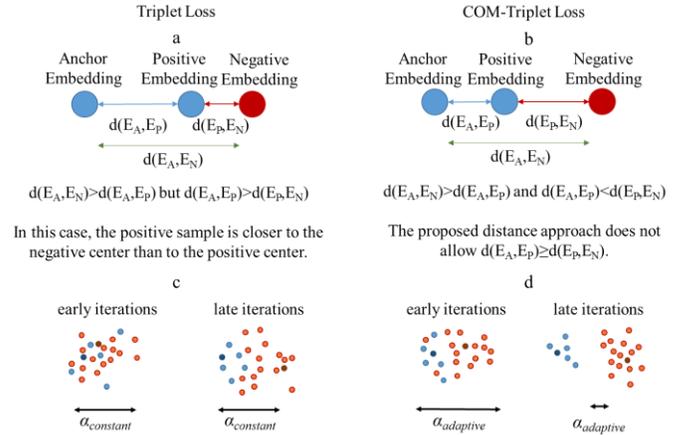

Figure 1. a) The traditional triplet loss forces the distance between $E_A$ and $E_P$ to be shorter than the distance between $E_A$ and $E_N$, but it does not consider the distance between $E_P$ and $E_N$. b) The proposed COM-Triplet loss considers all pair-wise distances among the image triplet, including the distance between $E_P$ and $E_N$. This ensures that all embedding vectors remain close to their cluster centers. c) The constant margin value used in traditional triplet loss may show large mismatch with the cluster separation during the course of training, yielding suboptimal performance. d) The COM-Triplet loss instead uses an adaptive margin value based on the distance between $E_P$ and $E_N$, to automatically adjust the margin value given cluster separation. (light red circles represent samples from class *N*, dark red circles represent cluster center of class *N*, light blue circles represent samples from class *P*, dark blue circles represent cluster center of class *P*)

To address these limitations, here we introduce a novel

COM-Triplet loss that instead aims to lower *A-P* distance relative to the average of *A-N* and *P-N* distances (Fig. 1b). Furthermore, a tuning-free margin value is introduced based on the *P-N* distance that adaptively changes with cluster separation (Fig. 1d). Accordingly, COM-Triplet loss is:

$$L_{COM-Triplet} = \frac{1}{M}\sum_{i=1}^{M}\max\left\{0, Dist_{wa}^i + \alpha_{adaptive}^i\right\} \quad (3)$$

where $Dist_{wa}$ represents within-cluster versus across-cluster distance and $\alpha_{adaptive}$ represents the adaptive margin value. $Dist_{wa}$ contains two opposing terms with the first attempting to reduce the within-cluster distance, whereas the second attempts to increase the across-cluster distance:

$$Dist_{wa}^i = d(E_A^i, E_P^i) - 0.5*(d(E_A^i, E_N^i) + d(E_P^i, E_N^i)) \quad (4)$$

and the adaptive margin value is set based on cluster separation:

$$\alpha_{adaptive}^i = 1 - d(E_P^i, E_N^i) \quad (5)$$

The discriminative embeddings are required to serve the eventual goal of melanoma detection, so cluster centers in *n*th iteration are computed for minority and majority classes as follows:

$$\begin{aligned}Cl_{maj}^n &= \frac{1}{2M}\sum_{i=1}^{M}(E_A^i + E_P^i) \\ Cl_{min}^n &= \frac{1}{M}\sum_{i=1}^{M}E_N^i\end{aligned} \quad n=1,2,\ldots,t \quad (6)$$

where *Cl* stands for a cluster center in the embedding space $Cl \in R^S$, *t* indicates the total number of iterations. Note that in each iteration, *A*, *P* and *N* triplet images are randomly re-selected from the training dataset. During iterative clustering, it is possible to get occasional updates with lower cluster separation than the preceding iteration. To ensure monotonously increasing cluster separation, the update to the cluster centers is performed as:

$$(Cl*_{min}, Cl*_{maj}) = \arg\max\left\{d(Cl_{min}^n, Cl_{maj}^n), d(Cl*_{min}, Cl*_{maj})\right\} \quad (7)$$

where undesirable updates are omitted, $Cl*_{min}$ and $Cl*_{maj}$ represent optimum values of $Cl_{min}$ and $Cl_{maj}$, $Cl* \in R^S$. The resultant cluster centers at the end of the training process serve as prototypes for the disease classes in the embedding space. Algorithm 1 outlines the training procedures for the proposed supervised deep clustering method, and Fig. 2 illustrates the overall model architecture.

---

**Algorithm 1** Pseudo-Code of SDC
**Input**: Dataset **X**
**Initialization:** *t*, parameters of *Φ*, **M, S**
while (***n<t***)
  Randomly select triplets for iteration $n, \{A^i, P^i, N^i\}_{i=1}^M$
  Create embeddings, $\{E_A^i, E_P^i, E_N^i\}_{i=1}^M = \phi\{A^i, P^i, N^i\}_{i=1}^M$
  Compute $D_{wa}^i$ and $\alpha_{adaptive}^i$ as in Eq.4 and Eq.5
  Compute $Cl_{\{min,maj\}}^n$ as in Eq.6
  Update cluster centers using Eq.7
  Update CNN parameters of *Φ* according to Eq. 3
**Outputs:** $Cl*_{min}$ and $Cl*_{maj}$

---

***Unsupervised Deep Clustering (UDC)***: We also introduce an unsupervised variant of deep clustering for cases where no label information is available in the training set, as illustrated in Fig. 2. Note that calculation of the triplet loss requires formation of an image triplet with *A-P* belonging to a first class and *N* belonging to the opposing class. To enable such categorization in the absence of external labels, we introduce a GMM module into the proposed architecture that generates image pseudo-labels.

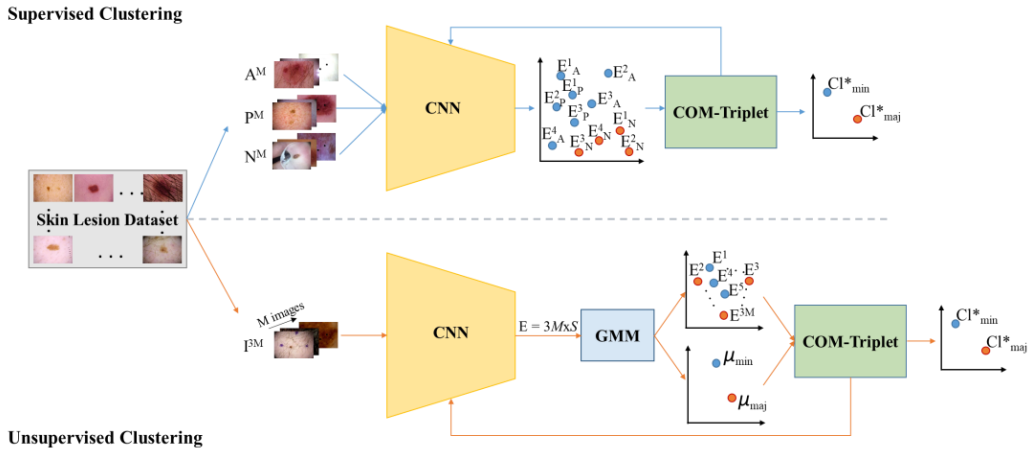

Figure 2. Deep clustering for melanoma detection. Top panel: For supervised clustering, M samples for A, P, and N images are randomly selected from the labeled dermoscopic dataset. The embedding vectors computed by the CNN module for each image triplet are used to define a COM-Triplet loss. The CNN parameters and the respective cluster centers for each class are learned based on this loss function. Bottom panel: For unsupervised clustering, a total of 3M images are randomly selected from the unlabeled dermoscopic dataset. Embedding vectors calculated by the CNN are assigned pseudo-labels via a GMM module, and these pseudo-labels are used to compute the COM-Triplet loss.

Accordingly, embeddings computed via the CNN module for a random batch of $3M$ images are probabilistically separated into two classes via the GMM module. The mixture model is expressed as a linear combination of multi-variate normal distributions in the embedding space:

$$f(E) = \sum_{k=1}^{2} h_k G_k(E; \mu_k, \Sigma_k) \quad (8)$$

where $h_k$, $\mu_k$, $\Sigma_k$ denote weight, mean, and covariance matrix, and $G_k$ is calculated as:

$$G_k(E; \mu_k, \Sigma_k) = \frac{1}{\sqrt{(2\pi)^S |\Sigma_k|}} e^{-0.5(E-\mu_k)^T \Sigma_k^{-1}(E-\mu_k)} \quad (9)$$

where $T$ indicates transpose, and $S$ denotes dimensionality of the embedding space. Assuming that unknown parameters for the entire GMM are aggregated as $\theta$, these parameters are identified by minimizing the negative log-likelihood of data samples under a positivity constraint for the mixture weights:

$$L_{NLL}(\theta) = \sum_{j=1}^{3M} \ln\left(\sum_{k=1}^{2} h_k G_k(E; \mu_k, \Sigma_k)\right) + \beta\left(\sum_{k=1}^{2} h_k - 1\right) \quad (10)$$

where $\beta$ is the Lagrange multiplier [50]. Once the GMM is trained, it can be used to assign each image sample to a Gaussian component to generate pseudo-labels:

$$r_{jk} = \frac{h_k G_k(E_j; \mu_k, \Sigma_k)}{\sum_{L=1}^{2} h_l G_l(E_j; \mu_l, \Sigma_l)} \quad (11)$$

where $r_{jk}$ is the membership probability of point $E_j$ to the $k$th component. Note that the number of samples from the minority class are expected to be low given the high degree of imbalance in the skin-lesion datasets (on average 1.7% in ISIC2020). In turn, a random batch of $3M$ images may contain only a small subset of the samples from the minority class that are repeatedly sampled thereby increasing risk of overfitting; or it may not contain any samples from the minority class at all. Therefore, we introduce a modified sample selection procedure for computing the triplet loss in the unsupervised scenario. In particular, the samples for $P$ and $N$ are replaced with their respective cluster centers. Each sample image considered as $A$ is first assigned to the closest cluster, and then its distances to its own cluster center versus the remaining cluster center are calculated. Furthermore, $\alpha_{adaptive}$ is also modified to compute the separation between positive and negative samples via the respective cluster centers. The resultant expressions for $\alpha_{adaptive}$ and $Dist_{wa}$ for unsupervised clustering are:

$$\alpha_{adaptive} = 1 - d(\mu_{min}, \mu_{maj}) \quad (12)$$

$$Dist_{wa}^i = \begin{cases} d(E_A^i, \mu_{min}) - 0.5*(d(E_A^i, \mu_{maj}) + d(\mu_{min}, \mu_{maj})), & \text{for } A^i \in C_{min} \\ d(E_A^i, \mu_{maj}) - 0.5*(d(E_A^i, \mu_{min}) + d(\mu_{min}, \mu_{maj})), & \text{for } A^i \in C_{maj} \end{cases} \quad (13)$$

***Inference Procedures***: At the end of model training, the proposed deep clustering method outputs two cluster prototypes for the minority and majority classes. To run inference on a test image, the CNN-based embedding of the input image is computed, and the distances of the image embedding from the two prototypes are characterized, $d_{maj}=d(Cl^*_{maj}, E_{Test})$, $d_{min}=d(Cl^*_{min}, E_{Test})$, where $d_{maj}$ represents the distance from the majority cluster center and $d_{min}$ represents the distance from the minority cluster center. Cluster assignment is performed based on the minimum of these distances:

$$Label = \begin{cases} C_{maj}, & \text{if } d_{maj} < d_{min} \\ C_{min}, & \text{if } d_{maj} > d_{min} \end{cases} \quad (14)$$

The inference procedure is illustrated in Figure 3, comprising the sequence of embedding generation, distance calculation to prototypes and cluster assignment.

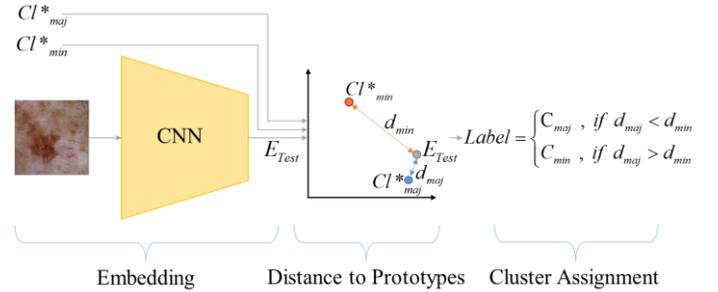

Figure 3. Inference procedure on a test skin-lesion image. The embedding vector of the test image is computed by the CNN module, and the distances of this embedding to the two cluster prototypes are then calculated. Cluster assignment is performed based on the minimum of the two distances.

## IV. EXPERIMENTS

***Dataset***: *The ISIC2020 dataset* contains 33126 dermoscopic images from 2056 patients [15, 51], with an average of 16 images per patient. Each image was assigned a benign (majority class) or malignant (minority class) label, and all melanoma cases were confirmed by histopathology whereas all benign cases were either reviewed by multiple experts or confirmed by histopathology. While 584 of the images in the ISIC2020 dataset contain melanoma, 32542 images are benign (minority class rate is 1.76%, and majority class rate is 98.24%). The skin-lesion images show a diverse set of resolutions ranging from 1872x1053 to 5184x3456 pixels. To avoid a complex CNN module and mitigate risks for overfitting, all images in the dataset were rescaled to a 256x256 spatial grid by performing a center-crop operation to subselect a square region containing the lesion followed by downsampling. The CNN module contained three channels to process RGB images.

*The ISIC2019 dataset* [42] contains 25331 training images from 8 different classes: melanoma, melanocytic nevus, basal cell carcinoma, actinic keratosis, benign keratosis, dermatofibroma, vascular lesions and squamous cell carcinoma. Only the melanoma and melanocytic nevus classes were used in this study. Accordingly, 4522 images in the melanoma class and 12875 images in the melanocytic nevus class were selected, and rescaled to a 256x256 spatial grid by performing a center-crop operation to subselect a square region containing the lesion followed by downsampling.

*The HAM10000 dataset* [41] consists of 10015 dermoscopic images in total. This dataset contains images of seven classes: melanoma, melanocytic nevus, basal cell carcinoma, actinic keratosis, benign keratosis, dermatofibroma and vascular lesions. Only the melanoma and melanocytic nevus classes were used in this study. Accordingly, 1113 images in the melanoma class and 6705 images in the melanocytic nevus class were selected, and rescaled to a 256x256 spatial grid by performing a center-crop operation to subselect a square region containing the lesion followed by downsampling.

***Architectural Details and Model Implementation***: The CNN module was designed based on common backbone architectures in computer vision tasks (VGG16 [52], ResNet50 [53], DenseNet169 [54], and EffcientNetB3 [55]). The dense layers of backbone CNNs were replaced with an embedding layer for deep clustering models. A dropout layer with a 0.3 dropout rate was added between the backbone CNN architectures and the embedding layer. The proposed model was implemented in Keras using the TensorFlow backend. All experiments were conducted on an NVIDIA RTX 3090 GPU. The Adam optimizer was used with $\beta_1$=0.9, $\beta_2$=0.99, $\varepsilon$=$10^{-7}$, learning rate $10^{-5}$, batch size 15, and number of epochs 15. In unsupervised clustering experiments, the number of mixture components was 2, the convergence threshold was $10^{-3}$, the non-negative regularization parameter for covariance was $10^{-6}$, and the k-means algorithm was used to initialize the GMM.

The deep clustering model was implemented with the proposed adaptive margin value, and ablated variants were trained using constant margin values of 0.2 or 0.6. These specific margin values were considered as they are most commonly reported in literature for the traditional triplet loss. Transfer learning and data augmentation were considered as learning strategies. For transfer learning, the backbone CNN weights were adopted from pre-trained models for object classification on the ImageNet database. For data augmentation, dermoscopic images were randomly shifted [-25, 25] pixels across the horizontal and vertical axes, flipped, rotated [-10 10] degrees, and/or scaled with a zoom factor of [90%, 110%].

To measure the training, validation, and test performances, each dataset was split into 75% training, 12.5% validation, and 12.5% test sets that did not overlap. The validation set was used to selected hyperparameters. When data augmentation was used, it was performed on the training set after the dataset split to prevent overlap. Performance was assessed by quantifying sensitivity, specificity, precision, accuracy, F1, and AUC metrics. Class-weighted averaging was used for metric calculations, as recommended in the Scikit libraries for imbalanced datasets.

To improve detection sensitivity during inference, feature selection was performed on the cluster prototypes $Cl^*_{min}$ and $Cl^*_{maj}$, where features with similar weights across the prototypes were neglected. Feature similarity was defined as an absolute difference of feature weights between the two prototypes that was lower than a threshold value. The threshold was taken as the difference between maximum and minimum feature weight averaged across prototypes.

Test performance on the ISIC2020 dataset was also measured by submitting the malignancy scores obtained on the officially released test set (which is different than the test set we obtained by three-way split of the official training set) to the ISIC challenge website https://challenge.isic-archive.com/. Note that the labels for the official test set are not publicly available, so we only reported AUC scores that were returned by the test site based on the input malignancy scores.

***Competing Supervised Methods***: We demonstrated the supervised variant of the proposed method (SDC) against direct CNN classifiers [16], [56] and deep clustering with traditional triplet loss [23]. Implementations of the competing methods are described below.

<u>Direct classifiers</u>: Direct classifier models were built based on the VGG16, ResNet50, DenseNet159, EfficientNetB3 backbone architectures. Input layers in each architecture were modified to receive 256x256x3 tensors for color images, and the output layers were modified with a softmax layer producing two outputs. Binary classification was performed based on binary cross-entropy loss. Training was performed via the Adam optimizer with learning rate $10^{-5}$, batch size 15, number of epochs 15. Several different learning strategies were considered including transfer learning, data augmentation and loss-term weighting. For transfer learning, direct classifiers that were pretrained for object classification on the ImageNet database were transferred to process skin-lesion images. Data augmentation procedures matched those used for deep supervised clustering. For loss-term weighting, the weighting procedure proposed in [56] was adopted where the weights were set inversely with the number of samples in the majority and minority classes in each iteration.

<u>Deep clustering</u>: Supervised deep clustering with the traditional triplet loss was implemented. A constant margin value of 0.2 was prescribed [23]. All other learning procedures were identical to that in SDC.

***Competing Unsupervised Methods***: In the absence of label information, we demonstrated the unsupervised variant (UDC) against shallow clustering, dimensionality reduction, decomposition and deep clustering methods. We considered GMM [50] and K-Means [57] as shallow clustering baselines, principal component analysis (PCA) [58], fast independent component analysis (Fast ICA) [59] and locally linear embedding (LLE) [60] as dimensionality reduction baselines, online dictionary learning (ODL) [61] as a decomposition baseline, and a convolutional autoencoder method (CAE) [62] and traditional triplet loss as deep clustering baselines. Implementations of the competing methods are described below.

<u>Shallow clustering:</u> A bivariate GMM was used with a convergence threshold of 0.001, and non-negative regularization for mixing weights was applied with parameter $10^{-6}$. A total of 100 expectation maximization iterations was performed. The k-means algorithm [57] was initiated with 10 different random seeds, and 3000 iterations were allowed. Trained cluster centers were used as in UDC for melanoma detection on test images.

*Dimensionality reduction*: For PCA, kernel principal components analysis based on a third-order linear kernel was employed. For FAST ICA, a fast implementation [59] was used with 200 iterations on whitened input data. For LLE, the neighborhood size was set as 5, the regularization constant was $10^{-3}$. Dermoscopic images were processed with a CNN backbone of matching architecture to that in UDC to compute embeddings; the image embeddings were then projected onto a single dimension via each dimensionality reduction method, and a threshold in this dimension was learned for classification.

*Decomposition*: For ODL, an orthogonal matching pursuit method was used. The sparsity parameter of the dictionary was 1, 1000 iterations were allowed. Dermoscopic images were processed with a CNN backbone of matching architecture to that in UDC to compute embeddings; the image embeddings were then projected onto a single dimension via ODL, and a threshold in this dimension was learned for classification.

*Deep clustering*: CAE was trained to reconstruct dermoscopic images from their noise-corrupted and randomly cropped versions. The feature vectors as computed by the encoder were processed with k-means to obtain two cluster centers. These centers were used as in UDC for melanoma detection on test images. Unsupervised deep clustering with the traditional triplet loss was also implemented. A constant margin value of 0.2 was prescribed [23]. All other learning procedures were identical to that in UDC.

## V. RESULTS

### A. Supervised Clustering for Melanoma Detection

To demonstrate the proposed approach, we first examined supervised deep clustering (SDC) for melanoma detection. We conducted a set of experiments to evaluate the influence of several important architectural and optimization parameters to the detection performance. These parameters included the backbone CNN (VGG16, ResNet50, DenseNet159, EfficientNetB3), margin value in triplet loss (adaptive versus constant values commonly reported in literature), and learning strategies (no pretraining, transfer learning, transfer learning and data augmentation). Performance metrics for variants of SDC are listed in Table I for the validation set, and in Table II for the test set. Among CNN backbones, VGG16 offers the highest performance with 3.14% improvement in AUC over the second-best variant. Using an adaptive margin value offers above 2.82% improvement over the constant margin values examined. Transfer learning by initializing the CNN backbone with weights pre-trained on the ImageNet database offers 3.34% higher AUC than a model trained on skin-lesion images with both transfer learning and data augmentation. This could be attributed to the repeated oversampling of the few minority class samples during data augmentation. The transfer learned model also offers 10.75% higher AUC than a model trained on skin-lesion images without any augmentation. These optimal configurations for SDC were used in all experiments thereafter.

TABLE I
VALIDATION PERFORMANCE OF SDC

| | | Recall | Precision | Specificity | Accuracy | F1-score | AUC |
|---|---|---|---|---|---|---|---|
| **Backbone CNNs** | *VGG16* | 97.46 | 97.72 | 98.52 | 97.46 | 97.59 | 93.30 |
| | *ResNet50* | 97.87 | 97.57 | 99.14 | 97.87 | 97.71 | 87.77 |
| | *DenseNet169* | 97.53 | 97.57 | 98.72 | 97.53 | 97.55 | 88.71 |
| | *EffcientNetB3* | 95.24 | 97.67 | 96.09 | 95.24 | 96.32 | 90.16 |
| **Margin** | $\alpha_{0.2}$ | 96.76 | 97.66 | 97.76 | 96.76 | 97.18 | 89.70 |
| | $\alpha_{0.6}$ | 97.32 | 97.62 | 98.43 | 97.32 | 97.46 | 90.48 |
| | $\alpha_{adaptive}$ | 97.46 | 97.72 | 98.52 | 97.46 | 97.59 | 93.30 |
| **Learning strategy (NP: no pretraining, TL:transfer learning, DA:data augmentation)** | NP | 91.98 | 97.44 | 92.85 | 91.98 | 94.45 | 82.55 |
| | TL | 97.46 | 97.72 | 98.52 | 97.46 | 97.59 | 93.30 |
| | TL+DA | 98.11 | 97.80 | 99.31 | 98.11 | 97.93 | 89.96 |

TABLE II
TEST PERFORMANCE OF SDC

| | | Recall | Precision | Specificity | Accuracy | F1-score | AUC |
|---|---|---|---|---|---|---|---|
| **Backbone CNNs** | *VGG16* | 96.95 | 97.03 | 98.40 | 96.96 | 96.99 | 88.11 |
| | *ResNet50* | 97.68 | 97.25 | 99.16 | 97.68 | 97.44 | 86.28 |
| | *DenseNet169* | 96.93 | 96.91 | 98.45 | 96.93 | 96.92 | 84.27 |
| | *EffcientNetB3* | 95.22 | 97.21 | 96.30 | 95.22 | 96.11 | 84.72 |
| **Margin** | $\alpha_{0.2}$ | 95.87 | 97.18 | 97.04 | 95.87 | 96.47 | 84.87 |
| | $\alpha_{0.6}$ | 96.54 | 96.94 | 97.98 | 96.54 | 96.74 | 86.82 |
| | $\alpha_{adaptive}$ | 96.95 | 97.03 | 98.40 | 96.96 | 96.99 | 88.11 |
| **Learning strategy (NP: no pretraining, TL:transfer learning, DA:data augmentation)** | NP | 92.22 | 97.05 | 93.22 | 92.24 | 94.38 | 84.06 |
| | TL | 96.95 | 97.03 | 98.40 | 96.96 | 96.99 | 88.11 |
| | TL+DA | 97.61 | 97.11 | 99.16 | 97.61 | 97.33 | 87.23 |

TABLE III
PERFORMANCE OF COMPETING METHODS ON THE ISIC2020 DATASET

| | | Split Test Set | | | | | | Official Test Dataset |
|---|---|---|---|---|---|---|---|---|
| | | Recall | Precision | Specificity | Accuracy | F1 | AUC | AUC |
| **Direct Classifiers (Backbone with transfer learning)** | *Classifier_{VGG16}* | 98.09 | 99.88 | 98.11 | 98.09 | 98.94 | 81.11 | 78.56 |
| | *Classifier_{ResNet50}* | 98.06 | 99.83 | 98.11 | 98.06 | 98.91 | 83.90 | 81.50 |
| | *Classifier_{DenseNet169}* | 97.17 | 98.25 | 98.02 | 97.17 | 97.77 | 82.78 | 82.72 |
| | *Classifier_{EfficientNetB3}* | 97.39 | 97.95 | 98.33 | 97.39 | 97.66 | 82.29 | 81.26 |
| **Direct Classifiers (NP: no pretraining, TL: transfer learning, DA:data augmentation, LW: loss weighting)** | *Classifier_{NP}* | 96.81 | 97.52 | 98.02 | 96.81 | 97.16 | 79.82 | 78.55 |
| | *Classifier_{TL}* | 98.06 | 99.83 | 98.11 | 98.06 | 98.91 | 83.90 | 81.50 |
| | *Classifier_{TL+DA}* | 98.02 | 99.95 | 98.04 | 98.02 | 98.97 | 83.76 | 82.54 |
| | *Classifier_{TL+LW}* | 96.35 | 94.47 | 98.67 | 96.35 | 95.41 | 84.23 | 82.52 |
| **Deep Clustering** | *SDC_{Triplet}* | 94.44 | 97.01 | 95.35 | 94.44 | 95.73 | 84.42 | 86.64 |
| | *SDC_{COM-Triplet}* | 96.95 | 97.03 | 98.40 | 96.96 | 96.99 | 88.81 | 88.89 |

To assess SDC against competing methods, detection performance was measured on the test set obtained via a three-way split of the ISIC2020 training data, and also on the official test dataset released with the ISIC challenge. Table III lists performance metrics for direct classifiers, clustering via the traditional triplet loss, and the proposed method. To examine the influence of backbone CNN, separate classifiers with different backbones were trained where network weights were initialized from models pre-trained on ImageNet (i.e., transfer learning). To examine the influence of learning strategy, classifiers with ResNet50 backbone with the highest validation performance were considered. Learning strategies included no pre-training, transfer learning, transfer learning and data augmentation, and transfer learning and loss-term weighting. Among direct classifiers, the model with ResNet50 backbone trained with transfer-learning and loss-term weighting yielded near-optimal performance in both test sets. Still, $SDC_{COM\text{-}Triplet}$ improves AUC by 4.58% over $Classifier_{TL+LW}$ in the split test set, and by 6.37% in the official test set. Furthermore, $SDC_{COM\text{-}Triplet}$ outperforms clustering with traditional triplet loss ($SDC_{Triplet}$) by 4.39% in the split test set, and by 2.25% in the official test set.

## B. Unsupervised Clustering for Melanoma Detection

Next, we examined the unsupervised deep clustering (UDC) for melanoma detection in cases where label information is absent in the training dataset. We again evaluated the influence of the backbone CNN, margin value in triplet loss, and learning strategies. Performance metrics for variants of UDC are listed in Table IV for the validation set, and in Table V for the test set. Among CNN backbones, VGG16 offers the highest performance with 4.2% improvement in AUC over the second-best variant. Using an adaptive margin value offers above 4.95% improvement over constant margin values examined. Transfer learning by initializing the CNN backbone with weights pre-trained on the ImageNet database offers 3.58% higher AUC than a model trained on skin-lesion images with both transfer learning and data augmentation, and 8.74% higher AUC than a model trained on skin-lesion images without any augmentation. These optimal configurations for UDC were used in all experiments thereafter.

To assess UDC against competing methods, detection performance was measured on split and official test sets in the ISIC2020 dataset. Table VI lists performance metrics for shallow clustering methods, dimensionality reduction methods, decomposition methods, and deep clustering based on auto-encoders, traditional triplet loss or COM-Triplet loss. Among competing methods, $UDC_{COM\text{-}Triplet}$ achieves the highest performance including deep clustering based on CAE and traditional triplet loss. $UDC_{COM\text{-}Triplet}$ improves AUC by 1.89% over the top contender CAE in the split test set, and by 2.48% in the official test set. Furthermore, in the unsupervised case, the benefits of COM-Triplet loss for deep clustering are more apparent over the traditional triplet loss. $UDC_{COM\text{-}Triplet}$ outperforms $UDC_{Triplet}$ by 12.32% in the split test set, and by 11.43% in the official test set.

TABLE IV
VALIDATION PERFORMANCE OF UDC

| | | Recall | Precision | Specificity | Accuracy | F1-score | AUC |
|---|---|---|---|---|---|---|---|
| Backbone CNNs | VGG16 | 98.35 | 96.74 | 99.99 | 98.35 | 97.54 | 76.85 |
| | ResNet50 | 94.54 | 96.74 | 96.07 | 94.54 | 95.62 | 68.72 |
| | DenseNet169 | 94.35 | 97.00 | 95.67 | 94.35 | 95.62 | 72.65 |
| | EffcientNetB3 | 97.77 | 96.73 | 99.41 | 97.78 | 97.25 | 68.25 |
| Margin | $\alpha_{0.2}$ | 97.12 | 94.98 | 99.99 | 97.12 | 96.05 | 66.11 |
| | $\alpha_{0.6}$ | 98.33 | 96.68 | 99.99 | 98.33 | 97.51 | 71.90 |
| | $\alpha_{adaptive}$ | 98.35 | 96.74 | 99.99 | 98.35 | 97.54 | 76.85 |
| Learning strategy (NP: no pretraining, TL:transfer learning, DA:data augmentation) | NP | 78.84 | 97.26 | 80.01 | 78.84 | 88.05 | 68.11 |
| | TL | 98.35 | 96.74 | 99.99 | 98.35 | 97.54 | 76.85 |
| | TL+DA | 98.31 | 96.25 | 99.99 | 98.31 | 97.28 | 73.27 |

TABLE V
TEST PERFORMANCE OF UDC

| | | Recall | Precision | Specificity | Accuracy | F1-score | AUC |
|---|---|---|---|---|---|---|---|
| Backbone CNNs | VGG16 | 98.04 | 96.12 | 99.99 | 98.04 | 97.07 | 70.85 |
| | ResNet50 | 94.30 | 96.15 | 96.11 | 94.30 | 95.21 | 66.43 |
| | DenseNet169 | 94.08 | 96.22 | 95.83 | 94.08 | 95.12 | 65.94 |
| | EffcientNetB3 | 97.41 | 96.11 | 99.36 | 97.41 | 96.76 | 63.11 |
| Margin | $\alpha_{0.2}$ | 96.24 | 95.46 | 99.99 | 96.24 | 95.85 | 58.89 |
| | $\alpha_{0.6}$ | 97.68 | 96.20 | 99.99 | 97.68 | 96.94 | 65.72 |
| | $\alpha_{adaptive}$ | 98.04 | 96.12 | 99.99 | 98.04 | 97.07 | 70.85 |
| Learning strategy (NP: no pretraining, TL:transfer learning, DA:data augmentation) | NP | 75.87 | 96.91 | 83.94 | 75.87 | 86.39 | 68.50 |
| | TL | 98.04 | 96.12 | 99.99 | 98.04 | 97.07 | 70.85 |
| | TL+DA | 97.86 | 96.02 | 99.99 | 97.86 | 96.94 | 69.16 |

TABLE VI
PERFORMANCE OF COMPETING METHODS ON THE ISIC2020 DATASET

| | | Split Test Set | | | | | | Official Test Dataset |
|---|---|---|---|---|---|---|---|---|
| | | Recall | Precision | Specificity | Accuracy | F1 | AUC | AUC |
| Shallow Clustering | GMM | 96.16 | 95.24 | 98.90 | 96.16 | 95.70 | 62.90 | 64.30 |
| | K-Means | 95.22 | 95.01 | 99.98 | 95.22 | 95.11 | 61.87 | 62.27 |
| Dimensionality Reduction | PCA | 95.71 | 96.29 | 99.99 | 95.71 | 96.00 | 63.59 | 61.82 |
| | Fast ICA | 94.34 | 96.07 | 99.99 | 94.34 | 95.21 | 62.47 | 61.67 |
| | LLE | 96.89 | 96.15 | 99.90 | 96.89 | 96.52 | 64.66 | 63.12 |
| Decomposition | ODL | 97.59 | 96.02 | 99.98 | 97.59 | 96.81 | 65.13 | 66.72 |
| Deep Clustering | CAE | 97.81 | 96.29 | 99.99 | 97.81 | 97.05 | 68.96 | 67.66 |
| | $UDC_{Triplet}$ | 96.77 | 94.92 | 99.39 | 96.77 | 95.84 | 58.53 | 58.71 |
| | $UDC_{COM\text{-}Triplet}$ | 98.04 | 96.12 | 99.99 | 98.04 | 97.07 | 70.85 | 70.14 |

### C. Analyses on Degree of Imbalance

A main motivation for deep clustering based on discriminative embeddings is to improve resilience against class imbalance. To systematically examine the effect of data imbalance, we compared the detection performance of SDC with direct classifiers while the degree of imbalance was systematically varied. In particular, we examined performance for six different sets with the following number of majority:minority samples: 4500:4500, 4500:2250, 4500:1125, 4500:300, 4500:125 and 4500:75, respectively. For this purpose, we used the malignant melanoma and benign nevus images in the ISIC2019 dataset. The number of nevus images was kept fixed at 4500, while the number of melanoma images was systematically changed. For a tightly controlled comparison, the same VGG16 backbone was adopted for both SDC and the direct classifier. Learning strategies were set to their optimal configurations for each method as reported in Section V.A.

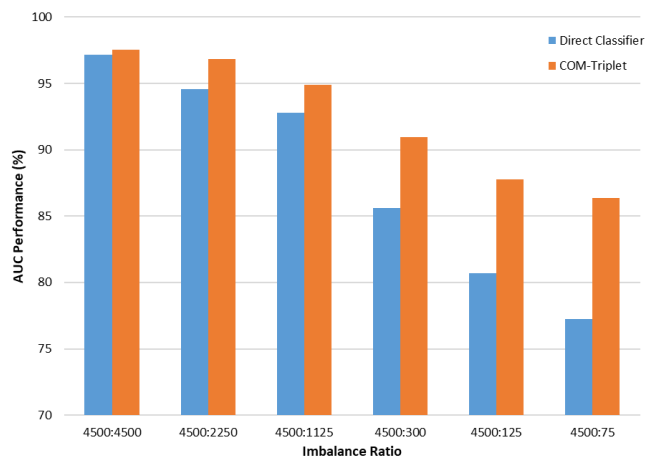

Figure 4. AUC performance of $Classifier_{VGG16}$ and $SDC_{COM\text{-}Triplet}$ under varying degrees of class imbalance simulated from the ISIC2019 dataset. The relative performance benefits of supervised deep clustering become more apparent towards higher imbalance ratios.

Figure 4 displays the AUC metrics for $SDC_{COM\text{-}Triplet}$ and $Classifier_{VGG16}$ trained separately on datasets with varying degrees of class imbalance. Naturally, performance for both methods is higher towards more balanced datasets. That said, performance of $Classifier_{VGG16}$ diminishes more rapidly towards higher imbalance ratios, whereas $SDC_{COM\text{-}Triplet}$ shows a more gradual decline in performance. As such, the benefits of $SDC_{COM\text{-}Triplet}$ over $Classifier_{VGG16}$ are most prominent in the highest imbalance ratios. While $SDC_{COM\text{-}Triplet}$ yields merely 0.35% higher AUC over $Classifier_{VGG16}$ at 4500:4500, it outperforms $Classifier_{VGG16}$ by 9.16% AUC at 4500:75.

### D. Analyses on Different Skin Lesion Datasets

Finally, we demonstrated the performance of the proposed approach on other public skin-lesion datasets. Specifically, we examined supervised and unsupervised melanoma detection in the popular ISIC2019 and HAM10000 datasets. It should be noted that the imbalance ratio in these two datasets for the melanoma class is notably limited when compared with the ISIC2020 dataset. While percentage difference between the occurrence rates of the classes (i.e., %nevus samples - %melanoma samples) is 96.48% in ISIC2020, it is 48.01% in ISIC2019 and 71.53% in HAM10000. As such, melanoma detection is a relatively easier task to implement on ISIC2019 and HAM10000 datasets.

TABLE VII
PERFORMANCE OF COMPETING METHODS ON ISIC2019 AND HAM10000 DATASETS

|  |  | ISIC2019 AUC | HAM10000 AUC |
|---|---|---|---|
| **Direct Classifiers** (Backbone with transfer learning) | $Classifier_{VGG16}$ | 96.33 | 97.71 |
|  | $Classifier_{ResNet50}$ | 97.05 | 97.90 |
|  | $Classifier_{DenseNet169}$ | 97.82 | 98.29 |
|  | $Classifier_{EfficientNetB3}$ | 97.60 | 98.16 |
| **Direct Classifiers** (NP: no pretraining, TL:transfer learning, DA:data augmentation, LW: loss weighting) | $Classifier_{NP}$ | 92.75 | 96.29 |
|  | $Classifier_{TL}$ | 97.82 | 98.29 |
|  | $Classifier_{TL+DA}$ | 97.70 | 98.30 |
|  | $Classifier_{TL+LW}$ | 98.08 | 98.62 |
| **Deep Clustering** | $SDC_{Triplet}$ | 97.85 | 98.46 |
|  | $SDC_{COM\text{-}Triplet}$ | 98.41 | 98.76 |

TABLE VIII
PERFORMANCE OF COMPETING METHODS ON ISIC2019 AND HAM10000 DATASETS

|  |  | ISIC2019 AUC | HAM10000 AUC |
|---|---|---|---|
| **Shallow Clustering** | GMM | 69.59 | 76.13 |
|  | K-Means | 67.88 | 75.45 |
| **Dimensionality Reduction** | PCA | 73.12 | 75.14 |
|  | Fast ICA | 72.82 | 74.62 |
|  | LLE | 76.27 | 76.63 |
| **Decomposition** | ODL | 74.57 | 76.03 |
| **Deep Clustering** | CAE | 79.89 | 77.82 |
|  | $UDC_{Triplet}$ | 76.44 | 75.67 |
|  | $UDC_{COM\text{-}Triplet}$ | 79.98 | 77.97 |

Table VII lists AUC for SDC and other competing methods in supervised settings. Compared with direct classifiers, $SDC_{COM\text{-}Triplet}$ has on average 1.21% higher AUC on ISIC2019 and 0.75% higher AUC on HAM10000. Moreover, $SDC_{COM\text{-}Triplet}$ outperforms $SDC_{Triplet}$ by 0.56% on ISIC2019 and 0.30% on HAM10000.

On the other hand, Table VIII lists AUC for UDC and other competing methods in unsupervised settings. Compared to shallow clustering methods, $UDC_{COM\text{-}Triplet}$ has on average 11.24% higher AUC on ISIC2019 and 2.18% higher AUC on HAM10000. Compared to dimensionality reduction methods, it has 5.91% higher AUC on ISIC2019 and 2.50% higher AUC on HAM10000. Compared to dictionary learning methods, it has 5.41% higher AUC on ISIC2019 and 1.94% higher AUC on HAM10000. Finally, compared against other deep clustering methods, $UDC_{COM\text{-}Triplet}$ offers 1.82% higher AUC on ISIC2019 and 1.23% higher AUC on HAM10000.

## VI. Discussion

Here we introduced a novel deep clustering approach for training melanoma detection models on highly imbalanced dermoscopy datasets. Since direct classifiers aim to maximize overall detection accuracy across classes, data samples from the minority class may have a limited effect on the trained model under heavy class imbalance. In contrast, our proposed method learns discriminative embeddings via a novel COM-Triplet loss that aims to maximize inter-cluster distances. Segregation of cluster centers in the embedding space is a proxy measure that is less susceptible to data imbalance between majority and minority classes compared to detection accuracy in classifiers [23, 63]. Pseudo-labels generated by a GMM module further enable unsupervised learning of cluster centers. Proximity to learned cluster centers is then used to detect skin lesions during inference.

Comprehensive demonstrations of both supervised and unsupervised variants of deep clustering were presented. Our experiments indicate that the proposed method outperforms direct classifiers and competing deep clustering methods in supervised settings, and shallow clustering, dimensionality reduction, decomposition, and competing deep clustering methods in unsupervised settings. Importantly, the proposed method shows improved reliability against data imbalance when compared to conventional classifiers. Furthermore, deep clustering via the proposed COM-Triplet loss outperforms that based on the traditional triplet loss in both supervised and unsupervised settings. Yet, the performance benefits are substantially higher for UDC. This pattern is most likely attributed to the use of an adaptive margin value in COM-Triplet as opposed to the fixed value in the traditional triplet loss. In the absence of label information for UDC, inter-cluster distances are expected to be relatively small in the early phases of the training procedure, so a constant margin value can yield suboptimal results. In contrast, the adaptive margin value in COM-Triplet can better accommodate the variability in the cluster estimates and their separation during the course of the training procedure.

Several prominent approaches have been considered in prior studies for model training on imbalanced datasets. The first group of methods resample imbalanced datasets to obtain relatively balanced numbers of samples from different classes. For instance, data augmentation or oversampling can be applied on the minority class, or undersampling can be performed on the majority class. While this strategy can be effective under moderate imbalance, it can increase the risk of overfitting by excessively oversampling the minority class under heavy imbalance such as that encountered in the skin lesion datasets considered here. An alternative group of methods instead perform pre-training in a data-abundant domain, and then fine-tune the models on balanced albeit compact skin lesion datasets. Although domain-transferred models partly mitigate the need for large training sets, they can show suboptimal performance when the pre-training and fine-tuning domains show divergent characteristics. In comparison, the proposed method performs training in the target domain, without resampling to balance the datasets.

In this study, we primarily focused on minimizing biases in trained models due to data imbalances between malignant and benign skin lesions. Yet, there are other aspects of modeling that can help improve task performance. A previous study has performed pre-processing for artifact removal in dermoscopy images to improve AUC in skin lesion detection [8]. Such pre-processing might also elicit performance improvements during deep clustering. Several prior studies have introduced ensemble learning for aggregating multiple classification models based on different CNN architectures for enhanced performance [45, 48]. When individual classifiers make non-overlapping prediction errors, ensemble models help boost overall performance. The proposed method might also benefit from ensemble learning with different CNN architectures in the backbone used to capture embeddings.

Here, we leveraged deep clustering based on COM-Triplet loss to detect melanoma in a two-class skin-lesion problem. As skin diseases show varying degrees of prevalence, training datasets for multiple skin diseases can possess similar multi-class imbalance problems. Thus, the proposed approach might also be useful in mitigating biases in multi-class detection of other skin diseases. The presented deep clustering approach might also be useful in the detection of other rare diseases based on dermoscopy such as nail [64] or hair disorders [65].

## VII. Conclusion

Melanoma is a rare disease compared to other causes of skin lesions, thus a native imbalance arises between malignant and benign samples in dermoscopic datasets. To alleviate biases due to class imbalance, here we presented a deep clustering method on discriminative embeddings learned via the COM-Triplet loss. Direct classification models tend to favor the majority class when trained on severely imbalanced datasets, even when data augmentation or transfer learning procedures are used. Instead, the proposed method produces maximally-separated cluster centers in a latent embedding space, where both majority and minority class samples contribute equally to distance calculations. We further show that the incorporation of a GMM module in the proposed architecture enables the generation of pseudo-labels for unsupervised training. Our results demonstrate that the proposed method outperforms several state-of-the-art baselines in both supervised and unsupervised setups. Therefore, it holds promise for improving reliability of deep-learning based melanoma detection.